# Surface smoothness requirements for the mirrors of the IXO X-ray telescope

D. Spiga[*], G. Pareschi

INAF/ Osservatorio Astronomico di Brera, Via E. Bianchi 46, I-23807 Merate (LC), Italy

**ABSTRACT**

The International X-ray Observatory (IXO) is a very ambitious mission, aimed at the X-ray observation of the early Universe. This makes IXO extremely demanding in terms of effective area and angular resolution. In particular, the HEW requirement below 10 keV is 5 arcsec Half-Energy Width (HEW). At higher photon energies, the HEW is expected to increase, and the angular resolution to be correspondingly degraded, due to the increasing relevance of the X-ray scattering off the reflecting surfaces. Therefore, the HEW up to 40 keV is required to be better than 30 arcsec, even though the IXO goal is to achieve an angular resolution as close as possible to 5 arcsec also at this energy. To this end, the roughness of the reflecting surfaces has to not exceed a tolerance, expressed in terms of a surface roughness PSD (Power-Spectral-Density). In this work we provide such tolerances by simulating the HEW scattering term for IXO, assuming a specific configuration for the optical module and different hypotheses on the PSD of mirrors.

## 1. INTRODUCTION

IXO[1], the International X-ray Observatory (launch foreseen in 2021), represents the evolution of the XEUS[2] X-ray telescope. IXO inherits from XEUS the main scientific targets[3], i.e., the observation of the early (z > 1) X-ray Universe, the study of the large-scale structure of the Universe, the galaxy formation, the measurement of Black Holes spin and accretion, the quest for missing baryons. Therefore, unprecedented optical performances are requested to IXO, like a very large effective area (3 $m^2$ at 1.25 keV, 0.65 $m^2$ at 6 keV, 150 $cm^2$ at 30 keV) and an excellent angular resolution: better than 5 arcsec below 10 keV with a goal of 2 arcsec.

To date, X-ray telescopes have been limited in their capabilities by the necessary trade-off between the angular resolution, the collecting area, the maximum size and the mass of the optics that can be operated in space environment. These different aspects are often in mutual competition. For example, it is well known that the ratio geometric area/mass increases with to the HEW of the mirror[4], therefore an improvement of the HEW has to be paid in terms of the attainable effective area. The technological challenge posed by IXO is currently being faced by an international collaboration (ESA, NASA, JAXA) devoted to the telescope development and to the assessment of technologies needed to fulfil the technical requirements. The huge effective area entails, as expected, a large mirror diameter (a few meters) and, because of the shallow reflection angles, a very long focal length. While XEUS was initially conceived with a 50 m focal length[2], later brought[3] to 35 m, for IXO a 20 m focal length is foreseen, as was to have been for SIMBOL-X[5], but using a single spacecraft, instead of two in formation flight. The distance between optics and detectors will be kept instead by means of an extendable bench[6].

The large size of the optics clearly rules out the possibility of manufacturing monolithic mirror shells, so they have to made of smaller segment that are assembled with the requested precision to return the desired optical performance. Moreover, an important point is represented by the mass of the optics, that should be limited to ~1800 kg[6]. This in turn requires the mirror segments to be made of lightweight material, like glass or Silicon. As a matter of fact, since 2004 the reference design of XEUS optics is based on *Silicon pore optics*[6],[7],[8]. Another possibility, also very promising, is based on the slumping of thin glasses[9],[10] presently carried out at NASA/GSFC and Columbia University for manufacturing the optics of NuSTAR[11].

Both glass and Silicon substrates of IXO optics need to be coated with X-ray reflective coatings, otherwise the effective area would be too small even in soft X-rays. To maximise the effective area at 0.1 to 10 keV, a single layer coating (Ir, Pt) would be suitable. Moreover, in order to enhance the reflectivity around 2-3 keV (where most high-Z elements have absorption edges) an overcoating with a low-Z material, like Carbon or Silicon Carbide, had been proposed[12] and proven to be effective[13]. Nevertheless, with the incidence angles envisaged for IXO, this kind of coating would leave a too small effective area around 30 keV, a spectral region where we expect to resolve the integrated accretion power of distant black holes, i.e., the X-ray background at its maximum. For this reason, it would be convenient to adopt graded multilayer coatings (W/Si) for mirrors with the smallest incidence angles to enhance their high-energy response.

---

[*] e-mail: daniele.spiga@brera.inaf.it

Another very important point, indeed, is the angular resolution. As we mentioned, IXO requires a HEW (*Half-Energy-Width*) of 5 arcsec or better, in the soft X-ray band (i.e., < 10 keV). In this energy range, the angular resolution is in general dominated by mirror figure deformations that can arise at manufacturing or mounting stages of the mirrors. As the photon energy increase, the *X-ray scattering* caused by the *surface roughness*[14] of mirrors also increases and can be the dominating factor for imaging degradation. Therefore, in order to ensure the optical performance over the energy band of sensitivity (i.e. < 40 keV), surface smoothness requirements for mirrors have to be clearly formulated, in addition to the figure accuracy that determines the HEW at all energies.

In this paper we define smoothness tolerances for the IXO optical module, on the basis of the HEW requirements (see Tab.1, compared with those of SIMBOL-X[5]). To do that, we might be making use of analytical formulae[15] derived from the well-known X-ray scattering theory[14], aimed at a direct translation of a HEW($\lambda$) requirement into a surface roughness PSD (*Power-Spectral-Density*). It is convenient, instead, to perform the inverse computation, by estimating the increase in the HEW from the known surface finishing of mirror substrates. The adopted method, already adopted in previous works[16],[17], is discussed in Sect. 2.

Tab. 1: comparison of the main characteristics of the IXO and SIMBOL-X optical modules

|  | IXO | SIMBOL-X |
|---|---|---|
| Focal length | 20 m | 20 m |
| Energy band | 0.1 – 40 keV | 0.5 – 80 keV |
| Effective area (1 keV) | 3 m$^2$ | 1400 cm$^2$ |
| Effective area (6 keV) | 0.65 m$^2$ (1 m$^2$ goal) | 600 cm$^2$ |
| Effective area (30 keV) | 150 cm$^2$ (350 cm$^2$ goal) | 450 cm$^2$ |
| Angular resolution (1 -10 keV) | < 5 arcsec HEW | 15 arcsec HEW |
| Angular resolution (10 – 40 keV) | 30 arcsec HEW | 20 arcsec HEW |

In Sect. 3 we present the result of the *on-axis* HEW computation for the IXO optical module. We hereafter consider only the case of Silicon pore optics because:

i) the surface of Silicon wafers is experimentally characterized[18] in terms of PSD, as it approximates well a power-law model[19] in a very wide range of spatial frequencies, *f*,

$$P(f) = \frac{K_n}{f^n},\qquad(1)$$

where the spectral index *n* is close to 1.7÷1.8 and $K_n$ takes on values in the range (2.0 ÷ 3.5) nm$^3$μm$^{-n}$, even though *these values cannot be always precisely determined*, because of errors in metrological measurements.

ii) for Silicon pore optics there is already an optical design[20] to fulfil the effective area requests at 1 and 6 keV. The optical design returns important pieces of information, useful to estimate the X-ray scattering contribution to the HEW; namely, the incidence angles on the mirrors, and the effective areas of mirrors segments with the same incidence angle (in the following referred to as "shells") to properly weight the contributions of the individual mirrors to the imaging degradation.

Finally, the results are briefly discussed in Sect. 4. We note that in this work we will not make any hypothesis regarding the origin of the 5 arcsec HEW. In fact, we simply assume that they result from the entire error budget of the optics manufacturing process (including the double cone approximation of the mirror profiles) at low X-ray energies, where the X-ray scattering is expectedly negligible. Our aim in this work is to demonstrate that, with the best achievable surface finishing of Silicon wafers, it is possible (though non trivial) to have an increase in the HEW due to scattering very low up to 10 keV, and largely within the specification up to 40 keV.

## 2. ADOPTED METHODOLOGY

**2.1. Imaging quality degradation factors**

The optical quality of the IXO mirrors is determined by essentially 3 factors;

1) *Diffraction and interference from pores' aperture*: this effect was extensively analyzed[21] because it represents the ultimate limit to the pore optics resolution. The impact on the angular resolution increases with $\lambda$, the wavelength of the radiation in use, with $f$, the focal length, and decreases with the pore size. For a 20 m focal length, the aperture diffraction effect appears to be important (i.e., ~ 1 arcsec) for energies below 1 keV[21] if the pores are in the range of 500-700 μm sizes.
2) *Mirrors and alignment deformations*: due to the composite nature of the IXO optical module, these are expected to arise at the integration/handling stage, more than at manufacturing. This kind of deformation is expected to cover the low-frequency part of the deformation spectrum; therefore the HEW degradation should be treated with the geometrical optic formalism, i.e., along with a ray-tracing routine. As such, the effect is usually expected to be independent of $\lambda$.
3) *X-ray Scattering (XRS) due to surface roughness:* this becomes dominant when the optical path differences of wavefronts reflected by different parts of the surface become comparable with $\lambda$. The relation between the surface roughness *Power Spectral Density* (PSD) and the scattered intensity distribution is well known from the scattering theory[14], and is also known to increase for increasing photon energy.

We are chiefly interested in the angular resolution in hard X-rays, i.e., beyond 10 keV; therefore we completely neglect the contribution of the aperture diffraction to the HEW degradation, while we discuss the effect of the geometrical deformations and the surface roughness. To this end, we assume that the effects (i.e., the HEW values) of geometry and of X-ray scattering can be added as

$$HEW^2(\lambda) \approx H_0^2 + H^2(\lambda). \qquad (2)$$

This approximation relies on the hypothesis that the two effects are independent of each other. The first term, $H_0$, is the contribution of the figure errors and the second, $H(\lambda)$, is due to the XRS.

As mentioned, the figure error term can be calculated once we know the actual shape of the mirrors, applying the geometrical optic and performing a detailed ray-tracing analysis. It is interesting to note that a coarse idea of the angular spread introduced by a mirror deformation spectrum can be retrieved from the slope rms[14] of the mirror, which in turn can be computed from the surface PSD, $P(f)$, as

$$\sigma_m^2 \approx \int_0^{f^*} (2\pi f)^2 P(f) df, \qquad (3)$$

where $f^*$ is the maximum frequency we can afford using in geometrical optics approximation. Then we suppose – crudely - that HEW ≈ FWHM = 2.35 $\sigma_m$ in Gaussian approximation, and we obtain

$$H_0 \approx 4.7 \left( 2 \int_0^{f^*} (2\pi f)^2 P(f) df \right)^{1/2}. \qquad (4)$$

The additional factor of 2 comes from the angle duplication due to reflection, and the factor √2 is due to the double reflection, supposing that the mirror deformations are uncorrelated for the two surfaces. We suppose the surface PSD to be a power-law $P(f) = K_n f^{-n}$ with $1 < n < 3$, as is often the case of optically-polished surfaces[18] like Silicon wafers. If the frequency $f^*$ were to go to infinity, the integral in Eqs. 3 and 4 would diverge even in the most optimistic case of $n = 3$. Fortunately, when $f^*$ becomes sufficiently large, the photon wavelength starts to be comparable to the size of surface defects, as projected along the incident direction of the ray. In such conditions, the concept of ray is no longer applicable, and physical optics is to be used to compute the surface diffraction (i.e., the XRS). Because of the proportionality of XRS diagram to the surface PSD, which in general decreases with $f$, the angular spread function of the mirror remains finite.

The boundary frequency, $f^*$, between geometrical and scattering treatment of optical defects has been investigated by Aschenbach[22] in terms of the rms of the Fourier components of rough profiles. On the basis of that criterion, figure prevails over scattering for all monochromatic components of roughness whose rms $\sigma_n$ fulfil the inequality

$$4\pi\sigma_n \sin\vartheta_i \gg \lambda. \tag{5}$$

In contrast, XRS dominates if $4\pi\sigma_n\sin\theta_i \leq \lambda$. On the other side, if we are reflecting with single layer coatings, a significant reflection occurs only for angles smaller than the critical one[23], $\sin\theta_i < (2\delta)^{1/2}$, where $\delta$ is 1 minus the real part of the coating refractive index. Hence, the condition for all wavelengths to fall in XRS regime is automatically satisfied if $4\pi\sigma_n(2\delta)^{1/2} \leq \lambda$. Considering the expression[23] of $\delta$, we obtain, finally,

$$\sigma_n < \sqrt{\frac{A}{16\pi N_A r_e \rho f_1}}. \tag{6}$$

Here $A$, $\rho$, $f_1$, are respectively the atomic weight, the density, the first scattering coefficient of the element used for the optical coating (e.g., Iridium), $N_A$ the Avogadro's number, $r_e = 2.8 \times 10^{-13}$ cm the classical electron radius. Note that Eq. 6 does not depend explicitly on either $\theta_i$ or $\lambda$. Moreover, we have that $f_1 \leq Z$, the atomic number of the coating material. Substitution of numerical values for Iridium yields a figure/scattering limit of a ~ 1 nm rms for each single component.

Now, assuming the power spectrum of Eq. 1, the integrated rms of *all* components reaches a 1 ÷ 2 nm rms (depending on the actual values used for $n$ and $K_n$) over the range of spatial wavelength shorter than 1 cm. We are then allowed to apply the XRS theory for almost all spatial frequencies, up to some cm. At such wavelengths one can expect that it is not the optical finishing of the Silicon wafer to be relevant, but the handling process. We then conclude that it makes sense to relate the X-ray scattering degradation to the roughness PSD of wafers, while the mirror figure remains independent of it, to a large extent. For this reason we assume in the remainder of this paper $H_0 = 5$ arcsec, whilst the $H(\lambda)$ is computed from a PSD of the kind reported in Eq. 1.

## 2.2. The HEW of a single mirror

Among a wealth of works describing method to relate the HEW degradation to the surface finishing of mirrors, we hereafter adopt an analytical method that allows us a fast computation of the scattering term of the HEW, without the need to trace the complete Point Spread Function of the mirror.

In an optical system with $N$ identical reflections at the incidence angle $\theta_i$, the scattering term $H(\lambda)$ can be directly computed from the mirror surface PSD $P(f)$, by solving for the spatial frequency $f_0$ the integral equation[15]

$$\int_{f_0}^{2/\lambda} P(f)df = \frac{\lambda^2}{16\pi^2 \sin^2\vartheta_i} \ln\left(\frac{2N}{2N-1}\right). \tag{7}$$

For mirror shells with Wolter-I or double cone profile, $N = 2$. Once $f_0$ is known, $H(\lambda)$ is calculated along with the first-order grating formula in the approximate form

$$H(\lambda) = \frac{2\lambda f_0}{\sin\vartheta_i}. \tag{8}$$

The application of the grating equation ensures that the optical path differences we are considering are of the order of $\lambda$, thus the scattering theory is still applicable. The total HEW is then calculated from Eq. 2. This approach is derived from the well-known Debye-Waller formula and offers the advantage to allow a quick computation of the HEW as a function of the photon energy. This method has also been confirmed numerically using the SIMBOL-X X-ray telescope as a test case[16]. There are, indeed, some limitations:

- the surface has to fulfil the smooth surface limit: this is required to justify the treatment of scattering in terms of surface diffraction (see Sect. 2.1)
- the focal length has to be long enough with respect to the mirror length for the incidence angle, $\theta_i$, to be constant along the axial profile;
- the scattering angles have to be small with respect to the incidence angle. This condition is fulfilled if the PSD is a steeply decreasing function, for the scattered energy at large angles to be of negligible intensity.

The last assumption can be weakened if we are computing the HEW over a limited field, as is the case in practice because a detector with a half-side (or radius) $r$ is able to collect only a part of the scattered X-rays. To account for this effect, the upper integration limit in Eq. 7 has to be replaced with the maximum observed spatial frequency $f_M$, corresponding to a scattering at $r$, i.e., $f_M = r \sin\theta_i/\lambda$.

If the field of view for the HEW computation is sufficiently large and the surface PSD is a power spectrum (Eq.1), the $H(\lambda)$ function can be computed analytically[15] as (here for $N = 2$)

$$H(\lambda) = 2\left[\frac{16\pi^2 K_n}{(n-1)\ln(4/3)}\right]^{\frac{1}{n-1}} \left(\frac{\sin\vartheta_i}{\lambda}\right)^{\frac{3-n}{n-1}}. \tag{9}$$

We see from this equation that the HEW scattering term increases for any $1 < n < 3$ with the photon energy and with the incidence angle through the $\sin\theta_i/\lambda$ ratio, to the $(3-n)/(n-1)$-th power. In hard X-rays (i.e., very small $\lambda$) it would then be convenient for the mirror to be characterized by a high spectral index, $n$, because this would slow down the increase in HEW for decreasing $\lambda$ (though at the expense of the HEW at low energies[15]).

Since for IXO we are mostly operating in total external reflection, for fixed $\lambda$ an upper limit to the incidence angle is the critical angle at the photon wavelength $\lambda$, i.e., $\theta_c \approx (2\delta)^{1/2}$ (see Sect. 2.1). If we recall the expression[23] for $\delta$, we find that an upper limit to $H(\lambda)$ is

$$H(\lambda) < C_n \left(\frac{N_A r_e \rho f_1}{\pi A} K_n^{\frac{2}{3-n}}\right)^{\frac{3-n}{2(n-1)}}, \tag{10}$$

where $C_n$ is a constant with respect to $\lambda$, which weakly depends solely on $n$. Substituting the numbers for Iridium we obtain that the expression in ( ) brackets is a pure number and, for typical values of $n$ and $K_n$, smaller than 1, therefore it *decreases* when its exponent *increases*. It is then convenient, *as long as $K_n$ is kept fixed* and if we use single Iridium layers to operate in total external reflection, to have mirror surfaces with a *small* $K_n$ and a *low* spectral index $n$, in order to make the HEW exponent *large* enough to minimize the right-hand side of Eq. 10. Probably, this would not be always true if multilayer coatings were used to extend the reflectivity beyond the critical angle, because Eq. 10 would no longer represent an upper limit to the HEW.

## 2.3. The HEW of the entire optical module

In the case we are considering, the HEW of every mirror segments of the IXO optics can be calculated along with Eq. 9, substituting the actual parameters ($n$ and $K_n$) of a Silicon wafer. Obviously the result will depend on $\theta_i$, i.e., the computed HEW($\lambda$) function is that of the mirrors "shell" (groups of mirrors pieces at the same distance $R$ from the axis, such that $R = f \tan(4\theta_i)$). The HEW$_T$ of the entire mirror module is expected to be a proper average of the contributions of the individual shells at different radii $R_k$, with incidence angles $\theta_{i,k}$ ($k = 1,2,\ldots M$ is the shell index, going inwards). It would then seem reasonable to average the individual HEW of the mirrors, HEW$_k$, over the respective *effective areas* $A_k$,

$$HEW_T^{(0)} = \frac{\sum_k HEW_k A_k}{\sum_k A_k}. \tag{11}$$

where we omitted the explicit dependence on $\lambda$ to simplify the notation. Indeed, we demonstrated[17] that such an average would lead to an overestimate of HEW$_T$. A more detailed approach, based on a first-order expansion of the $k^{th}$ PSF normalized to 1, $S_k$, around the $k^{th}$ half-energy radius, $\Omega_k$, suggests that the formula to be used is[17]

$$HEW_T^{(1)} = \frac{\sum_k HEW_k S_k(\Omega_k) A_k}{\sum_k S_k(\Omega_k) A_k}. \tag{12}$$

As in Eq. 11, we omitted the explicit dependence on $\lambda$. In particular, HEW$_T^{(0)}$ in Eq. 11 is obtained as an approximation of HEW$_T^{(1)}$ in Eq. 12 whenever one assumes that the variation of the coefficient $S_k(\Omega_k)$ with $k$ (i.e., with $\theta_{i,k}$) is negligible with respect to that of the effective areas. However, we verified, by means of a simple numerical example (see Fig. 1), that Eq. 12 is an underestimate, because the linear approximation of the PSF is not sufficiently accurate. It turns out, instead, that a very satisfactory approximation is given by the average of $HEW_T^{(0)}$ and $HEW_T^{(1)}$. In other words, the correct formula to be used to derive $HEW_T$ is

$$HEW_T^{(0)} = \frac{\sum_k HEW_k \cdot W_k}{\sum_k W_k}, \qquad (13)$$

where the weights, $W_k$, are expressed by

$$W_k = \frac{1 + S_k(\Omega_k)}{2} A_k. \qquad (14)$$

This correction was not accounted for in previous simulations[17] of the XEUS mirror module. This, in turn, caused an overestimate of the previously simulated HEW for XEUS.

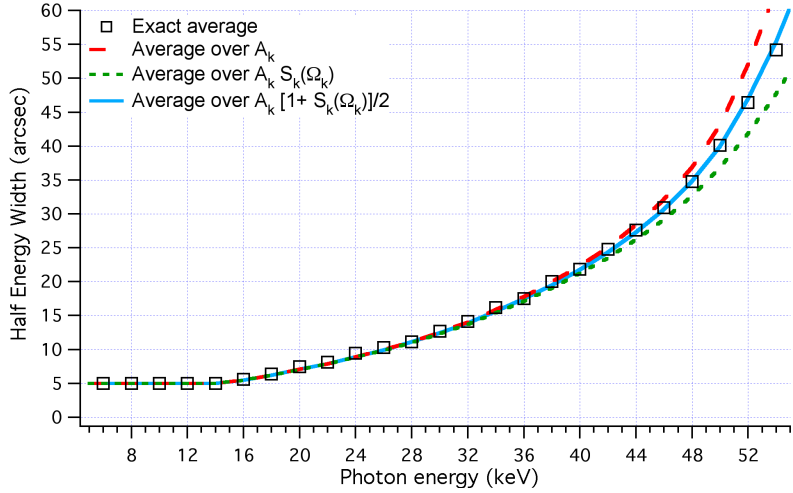

Fig. 1: numerical verification of the Eq. 14. The squares represent the exact HEW of a set of 12 mirror shells with incident angles in the range 0.13 - 0.2 deg, computed by summing PSFs with Lorentzian[14] shapes, $S(\theta) = 2/(\pi\Omega) \, [1+(\theta/\Omega)^2]^{-1}$, normalized to the respective effective areas, with $\Omega$ half-energy radius. The HEW values of the individual shells have been computed from a measured PSD, using Eq. 9, in a energy bandwidth from 5 to 55 keV, assuming a figure error $H_0 = 5$ arcsec, as foreseen for IXO. The HEW values were used in the exact calculation as parameter to simulate the PSF broadening with increasing energy due to XRS. The lines are the possible averages of the HEW values, as discussed in the text. The HEW average over the effective areas (dashed line) overestimates the correct HEW trend; accounting for the PSF values at the half-energy radius underestimates it (dotted line). The average of the two (solid line) allows reproducing much better the exact results.

## 3. ESTIMATION OF THE HEW OF IXO IN HARD X-RAYS

From the results reported in Sect. 2.2 it becomes apparent that the HEW scattering term increases as the ratio $\sin\theta_i / \lambda$ increases. This means that the outermost mirrors, for a fixed $\lambda$, suffer more from XRS than the smallest ones. On the other side, mirrors with larger $\theta_i$ have a lower critical energy for total reflection. Consequently, their contribution to the total effective area drops at a sufficiently large energy, then they are expected to have a lesser weight (Sect. 2.3) in the HEW degradation of the overall optical module of IXO.

The final HEW depends on which of the two effects prevail, which in turn depends on the mirror roughness PSD. If Silicon pores are adopted, the PSD to be considered is that of commercially available Silicon wafers (Eq. 1). Because of the variability/uncertainty in the parameters $n$ and $K_n$ characterizing the PSD, we have considered 4 possible PSDs crossing the two values of $n = 1.7, 1.8$ with those of $K_n = 2.0, 3.5$ nm$^3\mu$m$^{-n}$. The 4 different PSD are plotted in Fig. 2, and consecutively numbered with 1 to 4. In the same figure we also compare them with the PSD representing the roughness tolerance foreseen for the SIMBOL-X mirrors: this PSD tolerance was derived to minimize the XRS impact in hard X–rays, i.e., to have an HEW < 20 arcsec below 30 keV, 15 arcsec HEW of them due to figure errors. It can be seen that the surface finishing for SIMBOL-X mirrors is not far from the measured one of Silicon wafers. In Tab. 2 we also report some rms values in selected spectral bands, for the PSDs plotted in Fig. 2.

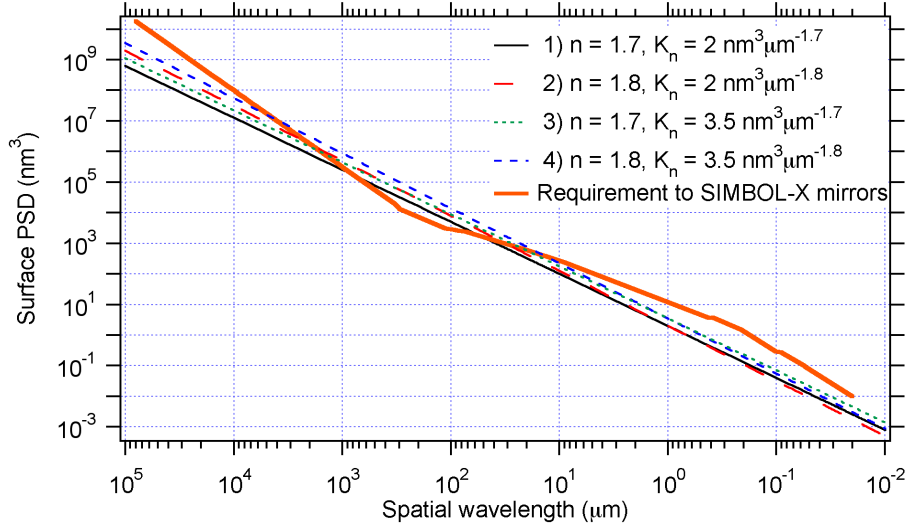

Fig. 2: different possible surface PSDs of Silicon wafers for the pore optics of IXO. Also the PSD tolerance foreseen for SIMBOL-X mirrors (relaxed with respect to a previous estimation[17], but still compliant with the SIMBOL-X HEW requirement) is displayed for comparison (see also Tab.2). Note that the spectra span over almost 7 decades.

For estimating the HEW degradation we may expect from the PSDs reported in Fig. 2, we need to know the angles at which X-rays are reflected, i.e., an optical design. In this respect, it has been recently demonstrated by R. Willingale[20] that the effective area requirements for IXO at 1 and 6 keV (see Tab.1) can be fulfilled using Silicon pore optics with a simple Iridium coating. In this design, the incidence angles, $\theta_{i,k}$, cover the range of 0.17 - 1.36 deg, and the distances from the optical axis, $R_k$, vary from 0.244 to 1.9 m, following the well-known relationship $R_k = f \cdot \tan(4\theta_{i,k})$. To reach the required effective area at 30 keV, finally, the mirror pieces with incidence angles smaller than 0.33 deg are supposed to be coated with graded multilayers[20].

We are not hereby discussing the details of the optical design for IXO mirrors: rather, we assume the range of radii and incidence angles given in[20] as an input (see Tab. 3) and a uniform radii/angles distribution in between. We can thereby compute, for each $k = 1,2…$ the effective area of the $k^{th}$ "shell", i.e., of all mirrors pieces at the same distance from the optical axis (the shell "radius"), and, consequently, of the same incidence angle,

$$A_k(\lambda) = 2\pi R_k \cdot \Delta R_k \cdot r^2(\lambda, \vartheta_{i,k}) \cdot V, \tag{15}$$

where $\Delta R_k$ is the constant wafer thickness (before being ribbed), $V$ the geometric efficiency of pores, $r(\lambda, \theta_{i,k})$ is the coating reflectivity at the photon wavelength $\lambda$ and the incidence angle $\theta_{i,k}$.

The resulting effective areas might be increased at low energies (~ 2 keV) by means of a Carbon overcoating[12],[13], but we do not account for this aspect in the present calculation. The effective area obstruction due to pore membrane and mechanical support structures, also discussed in detail in[20], is unessential for the present estimation as long as it can be considered independent of $k$. Also, we do not consider that only 7 of 8 sectors of the optical module are filled with mirrors for radii larger than 1.43 m, because the remaining sector is expected to not affect significantly the calculation.

Tab. 2: comparison of some roughness rms values in selected spectral bands, for the PSDs plotted in Fig. 2.

|  | 1 mm > $\ell$ > 100 µm | 100 µm > $\ell$ > 10 µm | 10 µm > $\ell$ > 1 µm | $\ell$ < 1 µm |
|---|---|---|---|---|
| SIMBOL-X requirement | 4.6 Å | 2.8 Å | 2.0 Å | 1.5 Å |
| 1) $n = 1.7$, $K_n = 2.0$ nm$^3$µm$^{-1.7}$ | 5.4 Å | 2.4 Å | 1.1 Å | 0.5 Å |
| 2) $n = 1.8$, $K_n = 2.0$ nm$^3$µm$^{-1.8}$ | 7.3 Å | 2.9 Å | 1.2 Å | 0.5 Å |
| 3) $n = 1.7$, $K_n = 3.5$ nm$^3$µm$^{-1.7}$ | 7.1 Å | 3.2 Å | 1.4 Å | 0.7 Å |
| 4) $n = 1.8$, $K_n = 3.5$ nm$^3$µm$^{-1.8}$ | 9.6 Å | 3.8 Å | 1.5 Å | 0.6 Å |

Tab. 3: some geometric parameters of possible designs of the IXO[20] and SIMBOL-X optical modules

|  | IXO (Silicon Pore optics) | SIMBOL-X |
|---|---|---|
| Focal length | 20 m | 20 m |
| Minimum radius | 0.244 m | 0.125 m |
| Maximum radius | 1.9 m | 0.321 m |
| Minimum incidence angle | 0.174 deg | 0.09 deg |
| Maximum incidence angle | 1.357 deg | 0.23 deg |

Since we are chiefly interested in the angular resolution at 30 keV, we need to compute the shell-by-shell effective area in the mirror module core with multilayer coatings, i.e. the set of mirrors with $\theta_{i,} < 0.33$ deg. For the IXO case we did not perform a multilayer stack(s) optimization: nevertheless, we note that (Tab. 3) the angle incidence range of IXO in which multilayers have to be used (0.17 – 0.33 deg) overlaps to that of SIMBOL-X, even if the minimum angle of SIMBOL-X is almost exactly ½ of the smallest incidence angle of IXO. On the other hand, IXO will be sensitive up to 40 keV, whilst SIMBOL-X was meant to have a bandwidth twice as large: for this reason, we have adopted one of the W/Si multilayer stack structure adopted in the SIMBOL-X development. This allowed us to compute the effective areas of the innermost shells of the IXO optical module and estimate their relative weight in the imaging degradation at 30 keV. As expected, the HEW scattering term at 30 keV is overwhelmingly dominated by their contribution.

We then computed the $H(\lambda)$ function for each shell, using Eq. 7 and Eq. 8, within a radius of 9 arcmin, corresponding to the field of view of IXO. The result did not significantly depend, indeed, on the particular field of view adopted and did not even sensitively differ from the Eq. 9. This was expected, indeed, from the quite steep power spectrum adopted. The HEW($\lambda$) for each shell is then derived from Eq. 2, assuming $H_0 = 5$ arcsec.

Finally, the HEW($\lambda$) should be averaged using the correct weights (Eq. 14). Nevertheless, since we do not know the detailed PSF of each mirror shell that accounts for both figure and scattering, we firstly neglected *any* figure error and performed the weighted average of the sole scattering terms, $H_k(\lambda)$. Therefore, the weights $W_k$ can be calculated by assuming the XRS diagram itself as PSF. The normalized XRS diagram at the photon wavelength $\lambda$, at the $k^{th}$ incidence angle, at a scattering angle equal to $\Omega_k = H_k(\lambda)/2$, returns the $S_k(\Omega_k)$ coefficients to be used in Eq. 14.

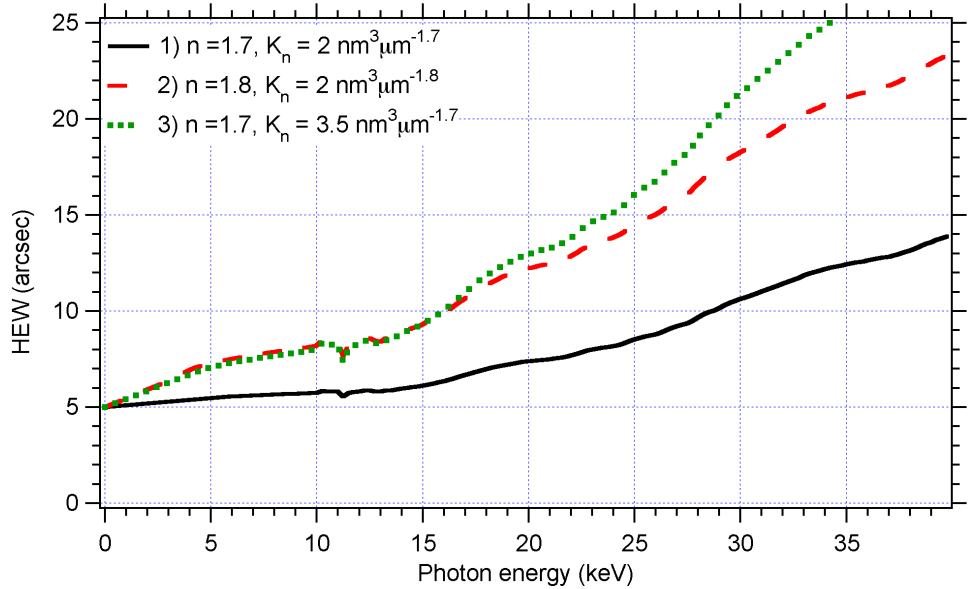

Fig. 3: the result of the HEW computation for 3 PSDs of Fig. 2, for the IXO optical module. The PSD n.1, with the lowest $n$ and $K_n$, returns the HEW curve (solid line) with the best performances. Increasing either $n$ (n.2, dashed line) or $K_n$ (n.3, dotted line) result in a performance worsening already in the soft X-ray band (< 10 keV). The requirement at < 40 keV is 30 arcsec.

Finally, the averaged $H(\lambda)$ term is added quadratically to the 5 arcsec HEW initially supposed for the geometric deformations and misalignments of the mirrors.

In Fig. 3 we report the final result of the computation for the first 3 PSDs displayed in Fig. 2 (the 4[th] would be out of specification at 40 keV). It can be seen that, as expected, the HEW of the IXO optical module critically depends on the surface PSD, at *low and high energies*. The results for the 3 considered cases are discussed hereafter.

1. *PSD No. 1*: $n = 1.7$ and $K_n = 2.0$ nm$^3$ μm$^{-1.7}$ (the solid line in Fig. 2). It is the most performing among the considered cases (Fig. 3). This was expected from the discussion of Sect. 2.2, and is a direct consequence of the values of $\sin\theta_i/\lambda$ of the IXO mirrors, limited by the critical angle for total reflection of Iridium. If this PSD is adopted, the IXO angular resolution requirements (see Tab.1) are fulfilled at low energies as the HEW increases only *slightly* in the energy band 1 – 10 keV (e.g., at 10 keV it is still less than 6 arcsec), whereas beyond 10 keV the HEW increases more rapidly, but still remains below 15 arcsec in the range 10 - 40 keV.
2. A steeper PSD like the *No. 2* ($n = 1.8$, $K_n = 2.0$ nm$^3$ μm$^{-1.8}$) causes the X-ray scattering to increase more rapidly even below 10 keV (the dashed lines in Fig. 2 and 3), up to 8 arcsec at 10 keV. This is a consequence of the larger roughness in the medium and low frequency part of the PSD ($\ell < 1$ μm). As the photon energy is increased, the angular resolution degrades more, but still within the nominal requirement at 40 keV (23 arcsec simulated Vs. 30 tolerable).
3. The *PSD No. 3* ($n = 1.7$, $K_n = 3.5$ nm$^3$ μm$^{-1.7}$, the dotted line in Fig. 2 and 3) is rougher than the PSD No. 1 at every frequency (see Tab. 2), although the spectral index is the same. It is, indeed, similar to the No. 2 in the low frequency regime (1 mm > $\ell$ > 100 μm) and rougher than the PSD No. 2 for $\ell$ < 100 μm. As a result, the HEW trends of the PSDs No. 2 and 3 are *very similar* up to 17 keV, then the PSD No. 3 gives rise to a larger increase in the HEW, that reaches 29 arcsec at 40 keV, just below the tolerance.

## 4. SUMMARY AND FUTURE DEVELOPMENTS

In this paper we presented a very preliminary computation of the HEW degradation due to XRS we can expect in the IXO telescope at 1 to 40 keV, adopting a proposed mirror configuration[20] based on Silicon pore optics, and assuming the typical roughness PSD of Silicon wafers, within the uncertainties of their surface finishing level. We have determined that small variations in the roughness have severe consequences on the HEW worsening for increasing photon energy. As long as the power-law parameters are close to $n = 1.7$ and $K_n = 2.0$ nm$^3$ μm$^{-1.7}$, the optical performances are expected to be fully compliant with IXO specifications (Fig.3), provided that a figure error HEW of 5 arcsec or less can be achieved. An increase in either $n$ or $K_n$ results in performance degradation, in soft and hard X-rays, but within reasonable limits the HEW should be smaller than 30 arcsec at 40 keV, therefore the high-energy HEW requirement (Tab.1) is probably too conservative. In contrast, the surface smoothness has to be very accurate, especially at spatial wavelengths larger than 10 μm, to meet the low-energy HEW requirement (see Tab. 2, the PSD No.1).

So far, the analysis was limited to the case of Silicon pore optics. Another possible technique to be adopted for manufacturing the IXO mirrors is the hot slumping[10] of thin glasses. Adoption of slumped glasses would in principle present the advantage that the surface roughness is determined, rather than the glass as-produced, by 1) the surface finishing level of the mould and 2) the capability of the glass to reproduce - in addition to its profile - the roughness of the mould, at least in the low and medium frequency range. Investigation of the surface roughness of hot-slumped glass sheets, spectral characterization and verification of their compliance with the IXO angular resolution specification will represent a development of the present work.

## ACKNOWLEDGMENTS

We thank R. Willingale (*University of Leicester*) and P. B. Reid (*Harvard-Smithsonian CfA*, Boston) for useful discussions.